\documentclass[a4paper,12pt]{article}

\usepackage[american]{babel}
\usepackage{csquotes}

\usepackage[margin=1.25in]{geometry}
\usepackage{setspace}
\usepackage{layouts}

\usepackage{graphicx}

\usepackage{float}
\usepackage{booktabs}
\usepackage{multicol}
\usepackage{multirow}
\usepackage[para,flushleft]{threeparttable}

\usepackage{algorithm}
\usepackage[noEnd=false,italicComments=false,beginComment=//~,beginLComment=/*~,endLComment=~*/]{algpseudocodex}

\usepackage{amsmath,amssymb,mathtools,bm}
\allowdisplaybreaks

\DeclareMathOperator{\sprct}{SSP-RCT}
\DeclareMathOperator{\drpt}{DRPT}
\DeclareMathOperator{\true}{T}
\DeclareMathOperator{\stated}{S}

\usepackage[]{amsthm}
\theoremstyle{plain}

\newtheorem{lemma}{Lemma}
\newtheorem{proposition}{Proposition}

\theoremstyle{definition}
\newtheorem{assumption}{Assumption}
\newtheorem{definition}{Definition}
\newtheorem{example}{Example}

\theoremstyle{remark}
\newtheorem{remark}{Remark}

\usepackage{enumitem}
\setlist[enumerate]{label=(\roman*)}
\newlist{propenum}{enumerate}{1} 
\setlist[propenum]{label=(\roman*), ref=\theproposition.(\roman*)}
\newlist{lemenum}{enumerate}{1} 
\setlist[lemenum]{label=(\roman*), ref=\thelemma.(\roman*)}
\newlist{assumptionenum}{enumerate}{1} 
\setlist[assumptionenum]{label=(\roman*), ref=\theassumption.(\roman*),nosep}

\usepackage{hyperref}
\hypersetup{colorlinks=true,allcolors=blue,bookmarksnumbered}

\usepackage[backend = biber,style = apa,sorting = nyt,sortcites = true,maxnames = 2,minnames = 1,hyperref = true,date = year,natbib = true,dashed = true,giveninits = true,uniquename=init]{biblatex}
\usepackage[capitalize,noabbrev,sort]{cleveref}
\crefname{theorem}{Theorem}{Theorems}
\Crefname{theorem}{Theorem}{Theorems}
\crefname{lemma}{Lemma}{Lemmas}
\Crefname{lemma}{Lemma}{Lemmas}
\crefname{corollary}{Corollary}{Corollaries}
\Crefname{corollary}{Corollary}{Corollaries}
\crefname{proposition}{Proposition}{Propositions}
\Crefname{proposition}{Proposition}{Propositions}
\crefname{assumption}{Assumption}{Assumptions}
\Crefname{assumption}{Assumption}{Assumptions}
\crefname{example}{Example}{Examples}
\Crefname{example}{Example}{Examples}
\crefname{equation}{}{}
\Crefname{equation}{}{}
\creflabelformat{equation}{(#2#1#3)}
\crefalias{propenumi}{prop}
\crefalias{lemenumi}{lem}
\crefalias{assumptionenumi}{assumption}

\title{Incorporating Preferences Into Treatment Assignment Problems%
    \thanks{I thank Takanori Ida for his helpful comments and continuous support.}
}
\author{Daido Kido%
    \thanks{Graduate School of Economics, Kyoto University. Email: \href{mailto:daido.kido@gmail.com}{daido.kido@gmail.com}}
}

\date{\today}

\begin{document}

\maketitle

\begin{abstract}
    This study investigates the problem of individualizing treatment allocations using stated preferences for treatments. If individuals know in advance how the assignment will be individualized based on their stated preferences, they may state false preferences. We derive an individualized treatment rule (ITR) that maximizes welfare when individuals strategically state their preferences. We also show that the optimal ITR is strategy-proof, that is, individuals do not have a strong incentive to lie even if they know the optimal ITR a priori. Constructing the optimal ITR requires information on the distribution of true preferences and the average treatment effect conditioned on true preferences. In practice, the information must be identified and estimated from the data. As true preferences are hidden information, the identification is not straightforward. We discuss two experimental designs that allow the identification: strictly strategy-proof randomized controlled trials and doubly randomized preference trials. Under the presumption that data comes from one of these experiments, we develop data-dependent procedures for determining ITR, that is, statistical treatment rules (STRs). The maximum regret of the proposed STRs converges to zero at a rate of the square root of the sample size. An empirical application demonstrates our proposed STRs.
\end{abstract}

\section{Introduction}\label{sec:introduction}
Preferences for treatments often affect their efficacy. Being assigned a disliked treatment makes an individual less motivated and less tolerant of any difficulties or inconveniences involved in that treatment. \textcite{Cook1979} termed such a phenomenon “resentful demoralization.” The presence of resentful demoralization possibly leads to the heterogeneity of the treatment effect with respect to the preference. This heterogeneity is sometimes called the \emph{preference effect}. The existence of the preference effect has been observed in several fields, including education \parencite{Wing2017,Little2008}, medical care \parencite{Long2008,Zoellner2019}, and energy saving programs \parencite{Ida2022}.
\par
In the presence of the preference effect, individualizing the treatment assignment for each true preference type is effective. If two treatments exist, say treatment 1 and 0, an example of the individualized assignments is the one giving treatment 0 to individuals preferring treatment 1 and giving treatment 1 to individuals preferring treatment 0. Such individualized treatment assignments based on individual characteristics (i.e., covariates) are called \emph{individualized treatment rule (ITR)}. Given any welfare function (typically, population mean outcome), the goal of individualization is to maximize welfare. The data-dependent decision of the ITR has been studied in the growing literature on statistical treatment choice \parencite[e.g.,][]{Manski2004,Athey2021,Kitagawa2018,Mbakop2021,Hirano2009}. 
\par
The literature usually assumes that the covariates used for individualization are \emph{observable} when an ITR is to be implemented. In the current study's context, this means that the true preference type is assumed to be observable. However, the \emph{true} preference is private information and \emph{unobservable} in nature. Instead, to implement the ITR, we must rely on the \emph{stated} preference by asking individuals. The true and stated preferences are not necessarily the same. On the contrary, when individuals are informed in advance about the ITR, some individuals have a strong incentive to tell a lie. For instance, recall the example of an ITR in the previous paragraph, that is, the ITR that gives the converse treatment to the preferred one. Consider individuals who prefer treatment 1 and suppose they know the ITR and are asked about their preference. Their truthful preference revelation gives them treatment 0, while the false preference revelation gives them the preferred treatment. Thus, telling a lie becomes the optimal behavior for these individuals. 
\par
Recently, several studies have analyzed individualized assignment problems with strategic agents \parencite{Munro2023,Sahoo2022,Harris2023}. In these problems, individuals have knowledge of the incoming ITR and strategically choose the values of their own covariates (not necessarily stated preference). Viewing stated preferences as a covariate, we can interpret our assignment problem (i.e., individualized assignment using stated preferences) as an instance of individualized assignment problems with strategic agents. Unfortunately, existing studies are not relevant to the current problem. This is because those studies have implicitly or explicitly assumed heterogeneous costs for choosing the difference level of covariate values. This presumption is adequate if ITRs use covariates such as test scores because improving test scores usually requires study effort. However, this presumption is inappropriate in the current problem as the preference statement is costless. Nevertheless, many real-world examples where the assignment of objects is determined based on the stated preference can be observed. A leading example is assignments of public schools to students \parencite[e.g.,][]{Abdulkadiroglu2005,Abdulkadiroglu2005a}. These assignments are based on the stated preference for schools. 
%
%
This study investigates the treatment assignment problem where ITRs use stated preferences and individuals know the applied ITR before the preference statement. First, we formally model the treatment assignment problem to individuals with preferences for treatments, which we briefly describe. There exist two treatments, treatment 1 and treatment 0, and each individual has a strict preference for treatments. The strictness excludes the indifference between distinct treatments. Hence, there exist two types of individuals in terms of preferences: individuals who strictly prefer treatment 1 and those who strictly prefer treatment 0.\footnote{The strictness of preferences is a common presumption in the literature of matching markets \parencite[e.g.,][]{Gale1962,Ergin2002,Roth1982}.} In this case, an ITR is described by the probability of giving treatment 1 for each preference type.\footnote{For simplicity, we do not consider covariates other than preference. The results of this study can accommodate covariates other than preferences as long as the additional covariates are discrete and not manipulatable.} In other words, an ITR is a pair of lotteries over treatments; one is given to individuals preferring treatment 1, and the other is given to individuals preferring treatment 0. The welfare function is set to the population mean outcome, following standard practice. The critical assumption on individuals’ preference statement is that each individual prefers the lottery that gives the preferred treatment with a higher probability. Equivalently, each individual maximizes their own expected utility, a standard assumption in microeconomics. 
\par
The first result, \cref{proposition:optimal-assignment-under-stated-preference}, gives the optimal ITR that maximizes welfare when individuals respond strategically to ITRs. The result leads to three findings. First, the knowledge of the \emph{true} preference type distribution and the conditional average treatment effect (CATE) given each \emph{true} preference type suffices for constructing the optimal ITR. The optimal ITR is determined by the signs of the CATEs multiplied by the share of the corresponding preference type. Second, the oracle ITR differs from the naive ITR that maximizes welfare while ignoring individuals' strategic preference statements. This suggests the significance of individuals' strategic behavior. Last, the optimal ITR is \emph{strategy-proof}, that is, no individual has a strong incentive to make a false preference statement. The strategy-proofness is regarded as a desirable property because strategy-proof ITRs reduce the burden of individuals’ thoughts. No matter how individuals contrive a scheme, there is nothing more to gain than to express their true preference. 
\par
To construct the optimal ITR, the distribution of the true preference type and the CATEs given the true preference type are necessary. This information is unknown in practice and must be identified and estimated from data. Unfortunately, however, the identification is not straightforward. For example, data in which individuals freely choose the preferred treatment is useless because no individual experiences the converse treatment to the preferred one. A naive idea that seems to work is to conduct a randomized controlled trial (RCT) with a pre-treatment survey on preference \parencite{Torgerson1996}. In this RCT, the pre-treatment survey first asks for the preferred treatment. Then, conditional on the answered preference type, treatments are randomly assigned. Unfortunately, this experiment does not necessarily identify the objects of interest, for the survey responses are stated preferences. The stated preferences do not necessarily correspond to true preferences unless the true preference revelation is adequately incentivized. For example, \textcite{Torgerson1996} conducted an RCT with the pre-treatment survey, where participants were informed that the treatments were randomly assigned with equal probability and asked about their treatment preferences. As the assignment probability did not depend on the stated preference, any preference statement was optimal behavior. As a result, the participants might have stated their preferences falsely. 
\par
To overcome the difficulty above, we introduce two particular experimental designs that allow us to identify the true preference type distribution and the CATE given the true preference type. One is the \emph{strictly strategy-proof RCT (SSP-RCT)}, an adjustment of the RCT with a pre-treatment survey so that the true preference revelation becomes the strictly optimal behavior for any individual. We demonstrate that under the assumption of individuals’ expected utility maximization, the SSP-RCTs can identify the objects of interest. The other experimental design is the \emph{doubly randomized preference trial (DRPT)} \parencite{Janevic2003,Rucker1989,Wennberg1993}. The DRPT randomly assigns individuals to treatment 0, treatment 1, and free-choice groups; in the former two groups, the treatment exposure is exogenously determined, while individuals’ choice determines it in the third group. As illustrated in \textcite{Ida2022,Wing2017,Long2008}, the DRPTs---combined with the exclusion restriction \parencite{Angrist1996}---can identify the target parameters. 
\par
Presuming data derived from data generating processes like the SSP-RCTs or DRPTs, we develop data-dependent procedures to determine an ITR, that is, \emph{statistical treatment rules (STRs)}. Following \textcite{Manski2004}, we evaluate the statistical performance of our proposed STRs with the maximum regret, the worst-case loss incurred due to not knowing the true data generating process. Specifically, as in \textcite{Kitagawa2018}, we derive finite-sample upper bounds of the maximum regret. These results imply that the maximum regret of the proposed STRs converges to zero at a rate of the square root of the sample size.
%
\paragraph{Related Literature}{This study contributes to the literature on individualized treatment assignment problems with strategic agents \parencite{Munro2023,Sahoo2022,Harris2023}. As mentioned above, the results of these studies do not apply to the problem this study addresses. This is because these studies focus on covariates that require some cost for manipulation, while the preference statement can be made without any cost. Moreover, the model has other differences. \textcite{Harris2023} consider a dynamic model while the deployed ITR is fixed over time. They assume that individuals have a homogeneous preference for treatments. \textcite{Sahoo2022} consider a dynamic model where the implemented ITR is consecutively updated and assumes a homogeneous treatment preference. Their model also incorporates capacity constraints to capture individuals' competition for scarce treatment. Unlike those two studies, this study develops a static model in which individuals have heterogeneous treatment preferences and no capacity constraints exist. The model of this study is very similar to that of \textcite{Munro2023}, the only difference being the cost of manipulating the covariates. See \cref{remark:model-comparison} for details.
\par
This study is also related to experimental designs incorporating individuals' preferences. \textcite{Brewin1989} propose an experimental design in which only individuals who are indifferent between treatments are randomly assigned; the other individuals are given the preferred treatment. \textcite{Zelen1990} proposes randomized consent designs where individuals are allowed not to comply with the randomly assigned treatment. The experimental designs proposed in \textcite{Rucker1989,Wennberg1993,Janevic2003} can be classified as DRPTs; that of \textcite{Wennberg1993} is most similar to the DRPT in this study. \textcite{Torgerson1996,Torgerson1998} discuss a combination of conventional RCTs with a pre-treatment survey on preferences. However, they do not consider the possibility that true and stated preferences may differ and thus do not discuss how to make individuals express true preferences. \textcite{Narita2021} also proposes a variant of RCT with a pre-treatment survey that achieves the Pareto efficiency among participants. However, the design is not totally strategy-proof but only approximately strategy-proof. The SSP-RCTs proposed in this study are versions of RCT with a survey that incentivizes truthful preference revelation at the expense of Pareto efficiency.
}
\paragraph{Organization of Paper}{The rest of this paper is organized as follows. \cref{sec:treatment-assignment-problems-with-preferences} formally models the individualized treatment assignment problem incorporating treatment preferences. We derive the optimal ITR that maximizes welfare under individuals' strategic preference revelation. \cref{sec:data-dependent-decision-of-itrs} discusses two experimental designs---the SSP-RCTs and DRPTs---that allow us to identify the distribution of true preference type and conditional average treatment effect given the true preference. Then, we propose the STRs associated with data from the SSP-RCTs and DRPTs. In addition, we evaluate the statistical performance of the proposed STRs with the maximum regret. \cref{sec:empirical-application} demonstrates the usefulness of the proposed STR using the results reported in \textcite{Wing2017}. \cref{sec:conclusion} concludes this paper. All proofs are relegated to \cref{sec:proofs}.
}

\section{Treatment Assignment Problems With Preferences}
\label{sec:treatment-assignment-problems-with-preferences}
This section models the individualized treatment assignment problem with treatment preferences and derives the optimal ITR. Specifically, \cref{sec:model} develops the model and \cref{sec:optimal-individualized-treatment-rules} discusses the optimal ITR that maximizes welfare.
\subsection{Model}
\label{sec:model}
We first discuss the standard model that assumes the observability of true preference. Then, we modify the model for the case when true preferences are not observable but stated preferences are observable.
\subsubsection*{Environment Under True Preference Observation}
Suppose two treatments exist, elements of $\mathcal{D}=\{0,1\}$. A policymaker plans to assign one of the treatments to each individual in the population of interest. The population is modeled as a probability space $(I,\Sigma,\mathbb{P})$, where $I$ denotes the set of individuals. For each treatment $d \in \mathcal{D}$, each $i \in I$ has a potential outcome $Y_i(d) \in \mathbb{R}$ that would be realized if $i$ was assigned treatment $d$. We maintain the stable unit treatment value assumption throughout the study.
\par
Suppose each individual has a \emph{strict} preference $\succsim_i$ for treatments (i.e., complete, transitive, and antisymmetric binary relation defined over $\mathcal{D}$). For treatments $d$ and $d'$, $d \succsim_i d'$ denotes that $i$ prefers $d$ to $d'$. The antisymmetric part of $\succsim_i$ is denoted by $\succ_i$. Note that the antisymmetricity rules out indifference between distinct treatments. Hence, there are only two types of individuals in terms of preferences: type 1 is those who strictly prefer 1, and type 0 is the converse. We denote the true preference type of individual $i$ by $T_i \in \mathcal{T} \coloneqq \{1,0\}$.
\par
The policymaker does not know the tuple $(Y_i(0), Y_i(1), T_i)$ for any individual $i$. Instead, suppose that the policymaker knows the joint distribution of $(Y_i(0), Y_i(1), T_i)$. Based on this information, the policymaker determines the probability of giving treatment 1 for each preference type. Formally, the policymaker chooses an \emph{individualized treatment rule (ITR)}, $\delta:\mathcal{T} \to [0,1]$, where $\delta(t)$ is the probability of giving treatment $t$ to individuals with preference type $t$.
\par
The standard treatment assignment problem \parencite[e.g.,][]{Manski2004, Kitagawa2018, Athey2021} assumes that all of the pre-treatment individual characteristics used for an ITR are \emph{observable} when the ITR is to be implemented. In the current setup, this means that the true preference type, $T_i$, is observable for any individual. Then, given an ITR $\delta$, the individual $i$'s treatment is drawn from the Bernoulli distribution with parameter $\delta(T_i)$. We refer to this setup as the \emph{environment under true preference observation}. The policymaker desires an ITR that maximizes the welfare defined as the expected outcome attained under an ITR. In the environment under true preference observation, given a joint distribution of $(Y_i(0), Y_i(1), T_i)$, the welfare under an ITR $\delta$ is
\begin{equation}
    W_{\true}(\delta) \coloneqq \mathbb{E}[Y_i(1) \delta(T_i) + Y_i(0) (1 - \delta(T_i))].
    \label{eq:def-welfare-true-preference}
\end{equation}
The law of iterated expectation yields
\begin{equation}
    W_{\true}(\delta) = \mathbb{P}(T_i = 1) \tau(1) \delta(1) + \mathbb{P}(T_i = 0) \tau(0) \delta(0) + \mathbb{E}[Y_i(0)],
    \label{eq:welfare-true-preference-expansion}
\end{equation}
where $\tau(t) \coloneqq \mathbb{E}[Y_i(1) - Y_i(0) | T_i = t]$ denotes the conditional average treatment effect (CATE) for true preference type $t$. From \cref{eq:welfare-true-preference-expansion}, we can easily describe the ITR that maximizes the welfare $W_{\true}$. Specifically, an ITR $\delta$ maximizes $W_{\true}$ if and only if it takes the form of 
\begin{equation}
    \begin{aligned}
        \delta(t) = 
        \begin{cases}
            1 & \text{if} \quad \mathbb{P}(T_i = t) \tau(t) > 0 \\
            \epsilon & \text{if} \quad \mathbb{P}(T_i = t) \tau(t) = 0\\
            0 & \text{if} \quad \mathbb{P}(T_i = t) \tau(t) < 0
        \end{cases}
        \quad\text{for each}\quad t \in \mathcal{T},
    \end{aligned}
    \label{eq:welfare-maximizing-ITR-true-preference}
\end{equation}
where $\epsilon \in [0,1]$. Namely, the welfare-maximizing ITRs are determined by the signs of CATEs weighted by the share of corresponding preference type. For later comparison, we refer to the ITR satisfying \cref{eq:welfare-maximizing-ITR-true-preference} as the \emph{naive} ITR.
\subsubsection*{Environment Under Stated Preference Observation}
We have formalized the treatment assignment problem, presuming the true preference is observable. Practically, the true preference type is an \emph{unobservable} feature. Instead, the policymaker must rely on the \emph{stated} preference type by asking each individual about their preferred treatment. Based on the stated preference type, the policymaker determines the treatment for each individual according to the prespecified ITR. Generally speaking, the true and stated preferences do not necessarily concur. On the contrary, when individuals know the ITR before the preference statement, some have a strong incentive to make a false preference statement as exemplified in \cref{example:manipulation-in-stated-preference}. 
\begin{example}\label{example:manipulation-in-stated-preference}
    Consider individuals who prefer treatment 1. The policymaker knows that the CATEs for each true preference type are $\tau(1) = -1$ and $\tau(0) = +1$. Then, the naive ITR $\delta$ under true preference observation is given by $\delta(1) = 0$ and $\delta(0) = 1$. Suppose the policymaker announces that this ITR will be implemented and asks individuals about their preference type. If the individuals state their true preference, then they are assigned treatment 0 certainly, but telling a lie gives them treatment 1 certainly. Hence, they have a strong incentive to make a false statement. Such a false statement leads to a significant welfare loss. Indeed, the individuals contribute to the welfare negatively since $\tau(1) = -1$.      
\end{example}
We explicitly distinguish between the true and stated preferences to discuss the welfare-maximizing ITR in the presence of individuals' strategic revelation of preference type. Let $S_i(\delta) \in \mathcal{T}$ be individual $i$'s stated preference type when $i$ knows the applied ITR is $\delta$. The stated preference type is allowed to differ from the true preference type. Note that the stated preference type is a function of ITRs, implying that the stated preference can differ depending on the ITR implemented. Then, the treatment assigned is drawn from the Bernoulli distribution with parameter $\delta(S_i(\delta))$. We refer to this circumstance as \emph{environment under stated preference observation}. As in the environment under true preference observation, the primal goal of the policymaker is to maximize the welfare (i.e., expected outcome). In the current environment, the welfare under an ITR $\delta$ is 
\begin{equation}
    W_{\stated}(\delta) \coloneqq \mathbb{E}[Y_i(1) \delta(S_i(\delta)) + Y_i(0) (1 - \delta(S_i(\delta)))].
    \label{eq:def-welfare-stated-preference}
\end{equation}
Comparing \cref{eq:def-welfare-true-preference,eq:def-welfare-stated-preference}, observe that $W_{\true}$ is modified by replacing the true preference, $T_i$, with the stated preference, $S_i(\delta)$. The two welfare functions generally disagree as $S_i(\delta)$ does not necessarily correspond to $T_i$. We refer to the ITR maximizing $W_S$ as the \emph{optimal} ITR.
\par
To proceed, we assume that the preference for treatments is naturally extended to the preference for lotteries over treatments (\cref{assumption:preference-for-lotteries}). This assumption allows us to describe when an individual makes the true preference statement. In its statement, a lottery over treatments is a vector $(p_1,p_0) \in [0,1]^2$ such that $p_1 + p_0 = 1$, where $p_d$ denotes the probability of getting treatment $d$.
\begin{assumption}[Preference for Lotteries Over Treatments]\label{assumption:preference-for-lotteries}
    For any two lotteries over treatments, $(p_1,p_0)$ and $(q_1,q_0)$, each individual $i$ [strictly] prefers $(p_1,p_0)$ to $(q_1,q_0)$ if and only if 
    \begin{equation}
        p_{T_i} \geq [>]~ q_{T_i}.
        \label{eq:first-order-stochastic-dominance}
    \end{equation}
\end{assumption}
This assumption says that each individual prefers the lottery that gives their preferred treatment with a higher probability. Hence, the preference for lotteries, characterized by \cref{eq:first-order-stochastic-dominance}, is a reasonable extension of the preference for treatments. We can easily observe that the condition \cref{eq:first-order-stochastic-dominance} holds if and only if $\mathbb{E}_{d \sim \mathrm{Ber}(p_1)}[u_i(d)] \geq \mathbb{E}_{d \sim \mathrm{Ber}(q_1)}[u_i(d)]$ for any utility function $u_i: \mathcal{D} \to \mathbb{R}$ representing $\succsim_i$\footnote{A real-valued function $u_i:\mathcal{D} \to \mathbb{R}$ is said to be a utility function representing $\succsim_i$ if and only if $d \succsim_i d'$ is equivalent to $u_i(d) \geq u_i(d')$ for any pair $(d,d')$ of treatments.}. Thus, an alternative interpretation of \cref{assumption:preference-for-lotteries} is that each individual maximizes their expected utility. This assumption is common in studies of matching markets \parencite[see, e.g.,][]{Kojima2010, Erdil2008, Erdil2014}. With a slight abuse of notation, we write $(p_1,p_0) \succsim_i [\succ_i] ~ (q_1,q_0)$ when individual $i$ [strictly] prefers $(p_1,p_0)$ to $(q_1,q_0)$. Note that two lotteries are the same if and only if any individual is indifferent between the two lotteries.
\par
Given the preference for lotteries over treatments, we can discuss whether an ITR incentivizes the true preference revelation. 
\begin{definition}\label{definition:strategy-proofness}
    An ITR $\delta$ is said to be \emph{$[strictly]$ strategy-proof} if each individual [strictly] prefers the lottery under the true preference statement to the lottery under the false preference statement; that is,
    \begin{equation*}
        (\delta(T_i),1-\delta(T_i)) \succsim_i [\succ_i] ~ (\delta(1-T_i),1-\delta(1-T_i))
    \end{equation*}
    for each $i \in I$.
\end{definition}
Under the strategy-proof ITR, each individual can obtain the lottery with (weakly) higher expected utility by telling the truth. In other words, any individual does not have a strong incentive to tell a lie in the preference statement. Moreover, the true preference revelation becomes the unique optimal behavior under the strictly strategy-proof ITRs. The following lemma gives a key to characterize strategy-proof ITRs.
\begin{lemma}\label{lemma:necessary-strategy-proofness}
    Suppose that \cref{assumption:preference-for-lotteries} holds. An ITR $\delta$ is [strictly] strategy-proof if and only if $\delta(1) \geq [>] ~ \delta(0)$. Moreover, the false preference revelation is the unique optimal behavior if and only if $\delta(1) < \delta(0)$.
\end{lemma}
\Cref{lemma:necessary-strategy-proofness} is helpful for checking whether an ITR is strategy-proof. An ITR is strategy-proof precisely when it gives treatment 1 to individuals whose stated preference type is 1 with a higher probability than individuals whose stated preference type is 0. The reason is apparent: because $\delta(1) \geq \delta(0)$, individuals preferring treatment 1 can get their preferred treatment with a higher probability by telling the truth. The inequality is equivalent to $1-\delta(0) \geq 1-\delta(1)$; thus, the above interpretation also holds for individuals desiring treatment 0.
\par
\cref{lemma:necessary-strategy-proofness} allows us to characterize the stated preference as follows:
\begin{equation}
    S_i(\delta) = 
    \begin{cases}
        T_i & \text{if} \quad \delta(1) > \delta(0),\\
        T_i\text{ or }1-T_i & \text{if} \quad \delta(1) = \delta(0),\\
        1-T_i & \text{if} \quad \delta(1) < \delta(0),
    \end{cases}
    \label{eq:stated-preference}
\end{equation}
for each $i$. That is, the true and stated preferences agree [disagree] for any individual when $\delta(1) > [<] ~ \delta(0)$. Note that the two lotteries under the true and stated preferences statements are the same when $\delta(1) = \delta(0)$. Therefore, all individuals are indifferent between the true and false preference statements. In this case, the stated preference can be arbitrarily chosen. The behavior described in \cref{eq:stated-preference} also yields a tractable representation of the welfare function in the environment under stated preference observation. Specifically, plugging \cref{eq:stated-preference} into \cref{eq:def-welfare-stated-preference} by cases, the welfare $W_{\stated}(\delta)$ under an ITR $\delta$ equals
\begin{align}
    &\begin{cases}
        \mathbb{E}[Y_i(1)\delta(T_i) + Y_i(0)(1-\delta(T_i))] & \text{if} \quad \delta(1) > \delta(0),\\
        \mathbb{E}[Y_i(1)\Bar{\delta} + Y_i(0)(1 - \Bar{\delta})] & \text{if} \quad \delta(1) = \delta(0) = \Bar{\delta},\\
        \mathbb{E}[Y_i(1)\delta(1-T_i) + Y_i(0)(1-\delta(1-T_i))] & \text{if} \quad \delta(1) < \delta(0).
    \end{cases}\nonumber\\
    = {} & 
    \begin{cases}
        \mathbb{P}(T_i = 1) \tau(1) \delta(1) + \mathbb{P}(T_i = 0) \tau(0) \delta(0) + \mathbb{E}[Y_i(0)] & \text{if} \quad \delta(1) > \delta(0),\\
        \mathbb{E}[Y_i(1)-Y_i(0)] \Bar{\delta} + \mathbb{E}[Y_i(0)] & \text{if} \quad \delta(1) = \delta(0) = \Bar{\delta},\\
        \mathbb{P}(T_i = 1) \tau(1) \delta(0) + \mathbb{P}(T_i = 0) \tau(0) \delta(1) + \mathbb{E}[Y_i(0)] & \text{if} \quad \delta(1) < \delta(0),
    \end{cases}
    \label{eq:welfare-stated-preference-expansion}
\end{align}
where equality follows from the law of iterated expectations. Note that the difference in the first and third cases in \cref{eq:welfare-stated-preference-expansion} is that the role of $\delta(1)$ and $\delta(0)$ are swapped. Comparison of the expansions of the two welfare functions given in \cref{eq:welfare-true-preference-expansion,eq:welfare-stated-preference-expansion} makes clear when the welfare functions in the environment under true and stated preference observation are different. The two welfare functions disagree when the false preference revelation is the unique optimal behavior.
\begin{remark}[Model Comparison]\label{remark:model-comparison}
    \textcite{Munro2023} discusses welfare-maximizing ITRs in situations where each individual strategically chooses the values of the features used in the ITRs. In particular, Theorem~1 in \textcite{Munro2023} postulates a sufficient condition for the welfare function $W_{\stated}(\delta)$ to be Gateaux differentiable at any point and characterizes the welfare-maximizing ITR. However, in the current model, $W_{\stated}(\delta)$ is not Gateaux differentiable at some point. For example, consider the ITR $\delta$ such that $(\delta(1),\delta(0))=(0.5,0.5)$, whence $W_{\stated}(\delta) = 0.5\mathbb{E}[Y_i(1) - Y_i(0)] + \mathbb{E}[Y_i(0)]$. The ITR moved by $\alpha$ in the direction of the ITR $h$ with $(h(1),h(0)) = (1,0)$ is characterized by $((\delta + \alpha h)(1),(\delta + \alpha h)(0)) = (0.5 + \alpha,0.5)$ for sufficiently small $\alpha$. When $\alpha > 0$, the true preference statement is the optimal behavior for any individual, whence $W_{\stated}(\delta+\alpha h) = 0.5\mathbb{E}[Y_i(1)-Y_i(0)] + \alpha\mathbb{P}(T_i=1) \tau(1) + \mathbb{E}[Y_i(0)]$. Conversely, when $\alpha < 0$, we have $W_{\stated}(\delta+\alpha h) = 0.5\mathbb{E}[Y_i(1)-Y_i(0)] + \alpha\mathbb{P}(T_i=0)\tau(0) + \mathbb{E}[Y_i(0)]$ because the false statement maximizes the expected utility for all individuals. Thus, $\lim_{\alpha \to 0} \alpha^{-1}(W_{\stated}(\delta + \alpha h) - W_{\stated}(\delta))$ does not exist in general, which implies the Gateaux differential of $W(\delta)$ at $\delta$ with increment $h$ does not exist. This is mainly because there exists no cost for false preference statements. Then, as an ITR, which incentivizes true preference revelation, moves so that false preference revelation becomes the strongly dominant strategy, a mass of individuals switch their strategy. As a result, the welfare function is not smooth enough to be directionally differentiable.
\end{remark}
\subsection{Optimal Individualized Treatment Rules}
\label{sec:optimal-individualized-treatment-rules}
As illustrated in \cref{example:manipulation-in-stated-preference}, the ITRs optimized ignoring individuals' strategic preference statements can lead to significant welfare losses. Then, the natural question is what kind of ITRs attain the highest welfare in the environment under stated preference observation. Moreover, are welfare maximization and strategy-proofness compatible? We answer these questions by deriving the oracle ITR under stated preference observation.
\begin{proposition}\label{proposition:optimal-assignment-under-stated-preference}
    Suppose that \cref{assumption:preference-for-lotteries} holds. The ITR $\delta^*$ given by 
    \begin{align}
        (\delta^*(1),\delta^*(0))
        = {} \begin{cases}
            (1,1) & \text{if} \quad \beta_1 > 0 \text{ and } \beta_0 > 0,\\
            (1,1) & \text{if} \quad \beta_1 > 0 \text{ and } \beta_0 = 0,\\
            (1,0) & \text{if} \quad \beta_1 > 0 \text{ and } \beta_0 < 0,\\
            (1,1) & \text{if} \quad \beta_1 = 0 \text{ and } \beta_0 > 0,\\
            (\epsilon,\epsilon) & \text{if} \quad \beta_1 = 0 \text{ and } \beta_0 = 0,\\
            (0,0) & \text{if} \quad \beta_1 = 0 \text{ and } \beta_0 < 0,\\
            (0,0) & \text{if} \quad \beta_1 < 0 \text{, } \beta_0 > 0 \text{, and } \beta_1 + \beta_0 < 0,\\
            (\epsilon,\epsilon) & \text{if} \quad \beta_1 < 0 \text{, } \beta_0 > 0 \text{, and } \beta_1 + \beta_0 = 0,\\
            (1,1) & \text{if} \quad \beta_1 < 0 \text{, } \beta_0 > 0 \text{, and } \beta_1 + \beta_0 > 0,\\
            (0,0) & \text{if} \quad \beta_1 < 0 \text{ and } \beta_0 = 0,\\
            (0,0) & \text{if} \quad \beta_1 < 0 \text{ and } \beta_0 < 0
        \end{cases}
        \label{eq:welfare-maximizing-ITR-stated-preference}
    \end{align}
    maximizes the welfare $W_{\stated}$ in the environment under stated preference observation for any joint distribution of $(Y_i(0), Y_i(1), T_i)$. Here, $\beta_1 \coloneqq \mathbb{P}(T_i = 1)\tau(1)$, $\beta_0 \coloneqq \mathbb{P}(T_i = 0)\tau(0)$, and $\epsilon \in [0,1]$ is arbitrary. Moreover, the ITR $\delta^*$ is always strategy-proof.
\end{proposition}
\begin{table}[t]
    \centering
    \begin{threeparttable}[t]
    \caption{Comparison of the Naive and Optimal ITRs}
    \label{tab:comparison-assignments}
    \begin{tabular*}{\textwidth}{@{\extracolsep{\fill}} ccccc}
        \toprule
        \multicolumn{3}{c}{Signs of Determinants} & \multicolumn{2}{c}{ITR}\\
        \cmidrule{1-3}\cmidrule{4-5}
        $\beta_1$ & $\beta_0$ & $\beta_1 + \beta_0$ & Naive & Optimal \\
        \midrule
        $>0$ & $>0$ & & $(1,1)$ & $(1,1)$\\
        $>0$ & $=0$ & & $(1,1)$ & $(1,1)$\\
        $>0$ & $<0$ & & $(1,0)$ & $(1,0)$\\
        $=0$ & $>0$ & & $(1,1)$ & $(1,1)$\\
        $=0$ & $=0$ & & $(\epsilon,\epsilon)$ & $(\epsilon,\epsilon)$\\
        $=0$ & $<0$ & & $(0,0)$ & $(0,0)$\\
        $<0$ & $>0$ & $< 0$ & $(0,1)$ & $(0,0)$\\
        $<0$ & $>0$ & $= 0$ & $(0,1)$ & $(\epsilon,\epsilon)$\\
        $<0$ & $>0$ & $>0$ & $(0,1)$ & $(1,1)$\\
        $<0$ & $=0$ & & $(0,0)$ & $(0,0)$\\
        $<0$ & $<0$ & & $(0,0)$ & $(0,0)$\\
        \bottomrule
    \end{tabular*}
    {\footnotesize
    \begin{tablenotes}
        \item\textit{Notes}: This table compares the naive and optimal ITRs. The first three columns show the signs of $\beta_1 = \mathbb{P}(T_i=1)\tau(1)$, $\beta_0 = \mathbb{P}(T_i=0)\tau(0)$, and $\beta_1 + \beta_0 = \mathbb{E}[Y_i(1)-Y_i(0)]$, where $T_i \in \{0,1\}$ denotes individual $i$'s true preference type; $T_i = 1$ if and only if individual $i$ strictly prefers treatment $1$ to treatment $0$. The conditional average treatment effect of individuals preferring treatment $t$ is denoted by $\tau(t)$; that is, $\tau(t) = \mathbb{E}[Y_i(1) - Y_i(0)|T_i = t]$. When the sign of $\beta_1 + \beta_0$ is implied by the signs of $\beta_1$ and $\beta_0$ or does not affect the oracle ITRs, the corresponding cell is left empty. The last two columns show the structure of the oracle ITRs under true and stated preference observation; for each cell, the first element is the probability of giving treatment 1 to individuals with true or stated preference type 1, and the second element is the probability of giving treatment 1 to individuals with true or stated preference 0. When $\beta_1 = \beta_0 = 0$ or $\beta_1 < 0 = \beta_1 + \beta_0 < \beta_0$, $\epsilon$ can be arbitrarily chosen from the unit interval.
    \end{tablenotes}}
    \end{threeparttable}
\end{table}
\cref{proposition:optimal-assignment-under-stated-preference} gives the optimal ITR $\delta^*$ under stated preference observation. This result yields three findings. First, the knowledge of $\beta_1 = \mathbb{P}(T_i=1)\tau(1)$, $\beta_0=\mathbb{P}(T_i=0)\tau(0)$, and $\beta_1 + \beta_0 = \mathbb{E}[Y_i(1) - Y_i(0)]$ are sufficient to construct the optimal ITR. In other words, it is sufficient to know the distribution of \emph{true} preference type and the CATEs given the \emph{true} preference type. The identification and estimation of the information will be discussed in \cref{sec:data-dependent-decision-of-itrs}.
\par
Second, the naive and optimal ITRs are different. To understand how individuals' strategic preference statements induce the difference, we construct \cref{tab:comparison-assignments}. \cref{tab:comparison-assignments} compares the naive and optimal ITRs given in \cref{eq:welfare-maximizing-ITR-true-preference,eq:welfare-maximizing-ITR-stated-preference}, by cases defined by the feature of the joint distribution of $(Y_i(0), Y_i(1), T_i)$. The first three columns show the signs of $\beta_1$, $\beta_0$, and $\beta_1 + \beta_0$. When the sign of $\beta_1 + \beta_0$ is implied by the signs of $\beta_1$ and $\beta_0$ or does not affect the structure of the oracle ITRs, the corresponding cell is left empty. To highlight the essential difference between the ITRs, the naive ITR is adjusted when its elements can be arbitrarily chosen from the unit interval to minimize the difference between the two ITRs. For instance, when $\beta_1 > 0$ and $\beta_0 = 0$, the naive ITR $\delta$ is given by $(\delta(1),\delta(0)) = (1,\eta)$ for arbitrary $\eta \in [0,1]$. In contrast, the optimal ITR is $(\delta^*(1),\delta^*(0)) = (1,1)$. In this case, we set $\eta = 1$ to make the two ITRs identical. When $\beta_1 = \beta_0 = 0$ or when $\beta_1 < 0 < \beta_0$ and $\beta_1 + \beta_0 = 0$, $\epsilon \in [0,1]$ can be arbitrarily chosen.
\par
Inspection of \cref{tab:comparison-assignments} reveals that the essential difference between the two ITRs exists precisely when $\beta_1 < 0$ and $\beta_0 > 0$. In this case, a policymaker who ignores individuals' strategic preference revelation will try to assign treatment $1$ only to individuals who genuinely prefer treatment $0$. However, each individual can gain by lying about their preferred treatment. As a result, the individuals receiving treatment 1 are precisely the opposite of those the policymaker originally aimed at. Instead, \cref{proposition:optimal-assignment-under-stated-preference} implies that assigning the same treatment uniformly to all individuals regardless of the stated preference type maximizes welfare. The uniform treatment is determined by the sign of the average treatment effect, $\beta_1 + \beta_0 = \mathbb{E}[Y_i(1)-Y_i(0)]$.
\par
Last, the optimal ITR is always strategy-proof: no individual has a strong incentive for false preference revelation under the optimal ITR. This is obvious from \cref{lemma:necessary-strategy-proofness}, since $\delta^*(1) \geq \delta^*(0)$ holds for any case. Moreover, $\delta^*(1)$ and $\delta^*(0)$ are equal except for the case when $\beta_1 > 0$ and $\beta_0 < 0$. In other words, individuals are indifferent between the two lotteries induced by the optimal ITR. Hence, individuals choose stated preferences arbitrarily. Nevertheless, this does not affect the welfare because the optimal ITR does not individualize the assignment. In contrast, the truthful preference revelation becomes the unique optimal behavior for all individuals when $\beta_1 > 0$ and $\beta_0 < 0$.

\section{Data-Dependent Decision of ITRs}
\label{sec:data-dependent-decision-of-itrs}
In \cref{sec:treatment-assignment-problems-with-preferences}, we assumed that the policymaker knows the distribution of the true preference type, $\mathbb{P}(T_i=1)$, and the average treatment effect conditional on the true preference type, $\tau(t) = \mathbb{E}[Y_i(1) - Y_i(0)| T_i = t]$. Practically, these objects are unknown and should be identified and estimated from data. In this section, we introduce two particular experiment designs that allow us to identify $\mathbb{P}(T_i=1)$ and $\tau(t)$. Specifically, \cref{sec:strategy-proof-randomized-controlled-trial} defines the \emph{strictly strategy-proof randomized controlled trial (SSP-RCT)}, an adjustment of the RCT with a pre-treatment survey, so that the true preference revelation becomes the strictly optimal behavior for any individual. \cref{sec:doubly-randomized-preference-trials} discusses the \emph{doubly randomized preference trial (DRPT)} \parencite{Rucker1989,Wennberg1993}. The DRPT randomly assigns individuals to treatment 0, treatment 1, and free-choice groups; in the former two groups, the treatment exposed is exogenously determined, while it is determined by individuals' choice in the third group. We demonstrate that both experimental designs can identify the objects of interest.
\par
Building on the identification of the key quantities, we develop data-dependent procedures to determine an ITR, presuming data derived from data generating processes like the SSP-RCTs or DRPTs. Concretely, we construct the \emph{statistical treatment rule (STR)}, a function that maps each possible realization of data to an ITR. 
Following \textcite{Manski2004}, we evaluate the performance of our proposed STRs based on the \emph{maximum regret}. Formally, given a class $\mathcal{P}$ of data generating processes and an STR $\widehat{\delta}$, the maximum regret of the STR is given by
\begin{equation*}
    \sup_{P \in \mathcal{P}} \mathbb{E}_{P}[\max_{\delta}W_{\stated}(\delta) - W_{\stated}(\widehat{\delta})].
\end{equation*}
The expectation corresponds to the regret, the average loss from the use of $\widehat{\delta}$ relative to the highest welfare achievable when the true data generating process $P$ is known. Then, the maximum regret is defined by taking the supremum of the regret over the class of the data generating processes. The class $\mathcal{P}$ will be specified below. We derive the finite-sample upper bound of the maximum regret of our proposed STR. These results imply that the worst-case regret converges to zero at rate $n^{-1/2}$.
\par
In the following analysis, we suppose that the sample population is the same as the population $(I,\Sigma,\mathbb{P})$ of interest.%
\footnote{Generally, the sample population can differ from the population of interest as long as the joint distribution of $(Y_i(1),Y_i(0),T_i)$ is the same between the two populations and individuals of the experimental population maximizes their own expected utility.} %
Thus, each member $i$ of the sample population has potential outcomes, $Y_i(0)$ and $Y_i(1)$, and the true preference type, $T_i$, and \cref{assumption:preference-for-lotteries} is satisfied.

\subsection{Strictly Strategy-Proof Randomized Controlled Trial}\label{sec:strategy-proof-randomized-controlled-trial}
An idea of the \emph{strictly strategy-proof randomized controlled trial (SSP-RCT)} is to adjust the propensity score of the RCT with a pre-treatment survey so that the true preference statement becomes the strictly optimal behavior for each individual. The trick to induce the true preference revelation comes from the observations in \cref{lemma:necessary-strategy-proofness}. To be specific, consider an propensity score function $p:\mathcal{T}\to[0,1]$ such that
\begin{equation}
    0 < p(0) < p(1) < 1
    \label{eq:strict-strategy-proofness}
\end{equation}
With this propensity score function being announced, each individual reports $S_i = S_i(p)$ in the pre-treatment survey. Then, each individual's experimental exposure $D_i \in \mathcal{D}$ is drawn from the Bernoulli distribution with parameter $p(S_i)$, and the outcome, $Y_i$, is observed according to $Y_i = Y_i(D_i)$. Thus, the observable data consists of $Y_i$, $D_i$, and $S_i$ for each $i$. Most importantly, condition \cref{eq:strict-strategy-proofness} ensures that the true preference statement becomes the utility-maximizing behavior (see \cref{lemma:necessary-strategy-proofness}). Therefore, we have $S_i = T_i$ for any individual $i$. In addition, the unconfoundedness holds by construction; that is, $(Y_i(0),Y_i(1)) \perp D_i \mid S_i$. As a result, this experimental design can identify $\mathbb{P}(T_i = 1)$ and $\tau(t)$. Specifically, it can be easily shown that
\begin{align}
    \begin{aligned}
    &\mathbb{P}(T_i = 1) = \mathbb{P}(S_i = 1)\\
    &\tau(t) =\mathbb{E}[Y_i|D_i = 1, S_i = t] - \mathbb{E}[Y_i|D_i = 0, S_i = t]
    \label{eq:SP-RCT-identification}        
    \end{aligned}
\end{align}
for any $t$ in the support of $T_i$.
\par
It is natural to ask whether observational studies containing stated preferences make the identification possible. The joint distributions of $(Y_i(0),Y_i(1),T_i,S_i,D_i)$ satisfying \cref{assumption:strictly-strategy-proof-RCT} are sufficient for the identification, given that the observable data consists of $Y_i = Y_i(D_i)$, $D_i$ and $S_i$.
\begin{assumption}\label{assumption:strictly-strategy-proof-RCT}
    The joint distribution of $(Y_i(0),Y_i(1),T_i,S_i,D_i)$ has the following properties:
    \begin{assumptionenum}
        \item\label{assumptionenum:bounded-outcome-SPRCT} (Bounded Outcome) There exists $M > 0$ such that $|Y_i(d)| \leq M$ for all $d \in \mathcal{D}$ and $i \in I$,
        \item\label{assumptionenum:strict-overlap-SPRCT} (Strict Overlap) There exists $\kappa \in (0,1/2)$ such that $\kappa \leq \mathbb{P}(D_i = 1|S_i = t) \leq 1-\kappa$ for all $t \in \mathcal{T}$,
        \item\label{assumptionenum:unconfoundedness-SPRCT} (Unconfoundedness) $(Y_i(0),Y_i(1)) \perp D_i \mid S_i$,
        \item\label{assumptionenum:agreement-SPRCT} (Agreement Between True and Stated Preferences) $S_i = T_i$ for all $i \in I$.
    \end{assumptionenum}
\end{assumption}
\cref{assumptionenum:bounded-outcome-SPRCT,assumptionenum:strict-overlap-SPRCT,assumptionenum:unconfoundedness-SPRCT} are standard in the study of statistical treatment rules \parencite[e.g.,][]{Kitagawa2018,Mbakop2021,Zhou2023}. \cref{assumptionenum:bounded-outcome-SPRCT} can be weakened to the existence of expectations, $\mathbb{E}[Y_i(1)]$ and $\mathbb{E}[Y_i(0)]$, for the identification. We include this assumption only for the regret analysis below. \cref{assumptionenum:agreement-SPRCT} requires that the true and stated preferences coincide for each individual. As illustrated above, this is satisfied if the joint distribution is induced by an SSP-RCT and \cref{assumption:preference-for-lotteries} holds. However, if we focus only on the satisfaction of \cref{assumptionenum:agreement-SPRCT}, this is possibly achieved by other methods. For instance, the literature on matching markets has developed strategy-proof assignment mechanisms \parencite[see, e.g.,][]{Roth1982,Dubins1981,Ergin2002}. For any distribution with \cref{assumption:strictly-strategy-proof-RCT}, the identification of $\mathbb{P}(T_i = 1)$ and $\tau(t)$ can be conducted in the same way as \cref{eq:SP-RCT-identification}. For fixed $M$ and $\kappa$, we denote by $\mathcal{P}_{\sprct}(M,\kappa)$ the class of joint distributions satisfying \cref{assumption:strictly-strategy-proof-RCT} because the SSP-RCTs particularly meet this assumption.
\par
Now, we propose the STR that maps the data generated from the joint distribution in $\mathcal{P}_{\sprct}(M,\kappa)$ to an ITR. Suppose that we obtain $n$ iid draws from the joint distribution of $(Y_i(0),Y_i(1),T_i,S_i,D_i)$ and observe data $\{(Y_i,D_i,S_i)\}_{i = 1}^n$, where $Y_i = Y_i(D_i)$. Given this data, $\mathbb{P}(T_i = t) \tau(t)$ can be unbiasedly estimated by 
\begin{equation}
    \widehat{\beta}_t = \frac{1}{n}\sum_{i=1}^n \frac{Y_i \cdot 1\{D_i = 1,S_i = t\}}{\mathbb{P}(D_i = 1|S_i = t)} - \frac{Y_i \cdot 1\{D_i = 0,S_i = t\}}{\mathbb{P}(D_i = 0|S_i =t)}.
    \label{eq:unbiased-estimator-SP-RCT}
\end{equation}
We assume that the propensity score, $\mathbb{P}(D_i=1|S_i=t)$, is known. Our proposed STR, $\widehat{\delta}_{\sprct}$, is defined by replacing $\beta_t$ in \cref{eq:welfare-maximizing-ITR-stated-preference} with $\widehat{\beta}_t$. For simplicity, we set $(\widehat{\delta}_{\sprct}(1),\widehat{\delta}_{\sprct}(0)) = (0,0)$ when $\widehat{\beta}_1 = \widehat{\beta}_0 = 0$ or when $\widehat{\beta}_1 < 0 < \widehat{\beta}_0$ and $\widehat{\beta}_1 + \widehat{\beta}_0 = 0$. The following result gives the statistical performance of our STR in terms of the maximum regret.
\begin{proposition}\label{proposition:maximum-regret-SP-RCT}
    Suppose that \cref{assumption:preference-for-lotteries,assumption:strictly-strategy-proof-RCT} hold. Then, for $n \geq \kappa^{-2}$,
    \begin{equation*}
        \sup_{P \in \mathcal{P}_{\sprct}(M,\kappa)} \mathbb{E}_{P^n}[\max_{\delta}W_{\stated}(\delta) - W_{\stated}(\widehat{\delta}_{\sprct})] \leq \frac{2e^{-1/2}M}{\kappa\sqrt{n}}.
    \end{equation*}
\end{proposition}
\cref{proposition:maximum-regret-SP-RCT} provides the finite-sample upper bound of the maximum regret of the proposed STR $\widehat{\delta}_{\sprct}$. Whatever joint distribution of $(Y_i(0),Y_i(1),T_i)$ the population has, the maximum regret of the STR converges to zero at rate $n^{-1/2}$ as long as the data comes from the data generating process meeting \cref{assumption:preference-for-lotteries,assumption:strictly-strategy-proof-RCT}. 
\begin{remark}[Comparison of Convergence Rate]
    \textcite{Kitagawa2018} develop an STR called \emph{empirical welfare maximization (EWM)}, presuming that individuals do not strategically respond to the ITR outputted from the EWM. They derive the finite-sample upper bound of the maximum regret of the EWM, which implies that the maximum regret converges to zero at rate $n^{-1/2}$. \cref{proposition:maximum-regret-SP-RCT} suggests that the convergence rate is identical to their result.
\end{remark}
\subsection{Doubly Randomized Preference Trials}\label{sec:doubly-randomized-preference-trials}
\emph{Doubly randomized preference trials (DRPTs)} randomly assign individuals to three experimental groups: treatment 0, treatment 1, and free choice groups \parencite{Rucker1989,Wennberg1993}. The exposed treatment is exogenously determined in the former two groups, and non-compliance is not allowed. Specifically, treatment $d$ is given in the treatment $d$ group. In contrast, each individual in the free-choice group freely chooses their preferred treatment. At first glance, the DRPT may seem a sole extension of the classical RCT with two treatment groups. However, the existence of the free-choice group, combined with an additional assumption, allows us to identify $\mathbb{P}(T_i = 1)$ and $\tau(t)$.
\par
We first introduce some variables to describe the DRPT formally. For ease of exposition, we denote the treatment 0, treatment 1, and choice group by 0, 1, and 2, respectively, and let $\mathcal{Z} = \{0,1,2\}$ be the set of the experimental groups. On top of $Y_i(0)$ and $Y_i(1)$, suppose that individual $i$ has a potential outcome $Y_i(d,z)$ that would be realized if $i$ were assigned to group $z$ and exposed to treatment $d$. For each $z \in \mathcal{Z}$, let $D_i(z) \in \mathcal{D}$ be the \emph{potential} treatment that individual $i$ would choose if $i$ was assigned to group $z$. As non-compliance is not allowed in the treatment 0 and 1 groups, we have $D_i(0) = 0$ and $D_i(1) = 1$ for all $i$. In the choice group, individuals choose the treatment according to their own preferences, whence $D_i(2) = T_i$ for each $i$. The DRPT determines the group to which $i$ belongs, $Z_i \in \mathcal{Z}$, by drawing a lottery over experimental groups. Then, the observable data consists of $Z_i$, the observed treatment $D_i = D_i(Z_i)$, and the observed outcome $Y_i = Y_i(D_i(Z_i),Z_i)$. By construction, the potential outcomes, potential treatments, and the true preference type are jointly independent of the assigned group; that is, $((Y_i(d,z))_{d\in\mathcal{D},z\in\mathcal{Z}},(D_i(z))_{z\in\mathcal{Z}}) \perp Z_i$.
\par
In DRPTs, the key assumption for the identification is the well-known exclusion restriction \parencite{Angrist1996}. That is,
\begin{equation}
    Y_i(d) = Y_i(d,z) \text{ for all } z \in \mathcal{Z} \text{ and for each } d \in \mathcal{D} \text{ and } i \in I. 
\end{equation}
This requires that whether the treatment exposed is determined exogenously or by their own choice does not affect the outcome. This assumption is often controversial in practice, but some methods exist to test its necessary condition. Specifically, \textcite{Kitagawa2015} provides a statistical test for the necessary condition of assumptions required to identify the local average treatment effect, that is, the random assignment of $Z_i$, monotonicity, and exclusion restriction \parencite{Angrist1996}. In DRPTs, the former two assumptions are automatically satisfied by construction, and hence, the procedure tests the necessary condition of the exclusion restriction. Alternatively, the discussion in Section~7 of \textcite{Long2008} suggests a test feasible under a particular experimental design that combines the strictly strategy-proof RCT and DRPT.
\par
Under the exclusion restriction, the DRPT can be used to identify $\mathbb{P}(T_i = 1)$ and $\tau(t)$ by viewing the assigned group $Z_i$ as the multi-valued instrumental variable \parencite{Ida2022,Wing2017}. First of all, $D_i(2) = T_i$ and random assignment of $Z_i$ implies 
\begin{align*}
    \mathbb{P}(T_i = 1) = \mathbb{P}(D_i = 1 | Z_i = 2).
\end{align*}
Because $D_i(2) = T_i \geq 0 = D_i(0)$, the CATE for individuals preferring treatment 1 is equivalent to the local average treatment effect (LATE) for individuals switching treatment as the instrument $z$ is exogenously changed from $0$ to $2$. More explicitly, we have
\begin{equation*}
    \tau(1) = \mathbb{E}[Y_i(1) - Y_i(0)|D_i(2) > D_i(0)].
\end{equation*}
Given this connection, the results in \textcite{Imbens1994,Angrist1996} imply that $\tau(1)$ can be identified as in
\begin{equation*}
    \tau(1) = \frac{\mathbb{E}[Y_i | Z_i = 2] - \mathbb{E}[Y_i | Z_i = 0]}{\mathbb{P}(D_i = 1 | Z_i = 2)}.
\end{equation*}
Similarly, the CATE for individuals preferring treatment 0 is the same as the LATE for individuals changing treatment as the exogenous switch of the instrument $z$ goes from $2$ to $1$. Hence, it follows that 
\begin{equation*}
    \tau(0) = \frac{\mathbb{E}[Y_i | Z_i = 1] - \mathbb{E}[Y_i | Z_i = 2]}{\mathbb{P}(D_i = 0 | Z_i = 2)}.
\end{equation*}
\par
With the interpretation of $Z_i$ as an instrument, \cref{assumption:DRPT} is sufficient for the identification of $\mathbb{P}(T_i = 1)$ and $\tau(t)$ using the instrumental variable approach described above.
\begin{assumption}\label{assumption:DRPT}
    The joint distribution of $((Y_i(d))_{d\in\mathcal{D}},(Y_i(d,z))_{d\in\mathcal{D},z\in\mathcal{Z}},(D_i(z))_{z\in\mathcal{Z}},T_i,Z_i)$ has the following properties:
    \begin{assumptionenum}
        \item\label{assumptionenum:bounded-outcome-DRPT} (Bounded Outcome) There exists $M > 0$ such that $|Y_i(d)| \leq M$ for all $d \in \mathcal{D}$ and $i \in I$,
        \item\label{assumptionenum:strict-overlap-DRPT} (Strict Overlap) There exists $\kappa \in (0,1/2)$ such that $\kappa \leq \mathbb{P}(Z_i = z) \leq 1-\kappa$ for all $z \in \mathcal{Z}$,
        \item\label{assumptionenum:random-assignment-DRPT} (Random Assignment) $((Y_i(d,z))_{d\in\mathcal{D},z\in\mathcal{Z}},(D_i(z))_{z\in\mathcal{Z}}) \perp Z_i$,
        \item\label{assumptionenum:existence-exogeneous-free-choice} (Existence of Exogeneous and Free Choices of Treatment) $D_i(0) = 0$, $D_i(1) = 1$, and $D_i(2) = T_i$ for all $i \in I$.
        \item\label{assumptionenum:exclusion-restriction-DRPT} (Exclusion Restriction) $Y_i(d) = Y_i(d,z)$ for all $z \in \mathcal{Z}$ and for each $d \in \mathcal{D}$ and $i \in I$.
    \end{assumptionenum}
\end{assumption}
\cref{assumptionenum:bounded-outcome-DRPT,assumptionenum:strict-overlap-DRPT,assumptionenum:random-assignment-DRPT} are parallel to \cref{assumptionenum:bounded-outcome-SPRCT,assumptionenum:strict-overlap-SPRCT,assumptionenum:unconfoundedness-SPRCT} in \cref{sec:strategy-proof-randomized-controlled-trial}. \cref{assumptionenum:existence-exogeneous-free-choice} requires that the instrument creates groups under which the exposed treatments are determined exogenously and a group in which individuals freely choose according to their preference. The joint distribution induced by the DRPT fulfills this requirement. Under the joint distribution satisfying \cref{assumption:DRPT}, $\mathbb{P}(T_i = 1)$ and $\tau(t)$ are identified in the same manner as above. We denote the class of joint distributions with \cref{assumption:DRPT} by $\mathcal{P}_{\drpt}(M,\kappa)$ for fixed $M$ and $\kappa$ because DRPTs satisfy the assumption.
\par
We now propose the STR mapping the data generated from the data generating processes satisfying \cref{assumption:DRPT} to an ITR. Suppose that we obtain $n$ iid draws from the joint distribution in $\mathcal{P}_{\drpt}(M,\kappa)$ and observe data, $\{(Y_i,D_i,Z_i)\}_{i = 1}^n$ following $D_i = D_i(Z_i)$ and $Y_i = Y_i(D_i(Z_i),Z_i)$. Given this data, one can unbiasedly estimate $\beta_t = \mathbb{P}(T_i = t) \tau(t),t\in\mathcal{T}$ by 
\begin{equation}
    \begin{aligned}
        \widehat{\beta}_1 &= \frac{1}{n}\sum_{j = 1}^n \frac{Y_i \cdot 1\{Z_i = 2\}}{\mathbb{P}(Z_i = 2)} - \frac{Y_i \cdot 1\{Z_i = 0\}}{\mathbb{P}(Z_i = 0)}\\
        \widehat{\beta}_0 &= \frac{1}{n}\sum_{j = 1}^n \frac{Y_i \cdot 1\{Z_i = 1\}}{\mathbb{P}(Z_i = 1)} - \frac{Y_i \cdot 1\{Z_i = 2\}}{\mathbb{P}(Z_i = 2)}
    \end{aligned}
\end{equation}
Again, the probabilities of group assignment, $\mathbb{P}(Z_i = z),z\in\mathcal{Z}$, are assumed to be known. This supposition is reasonable when the data is obtained from the DRPT. One can view $\widehat{\beta}_1$ and $\widehat{\beta}_0$ as unbiased estimators for the intention-to-treat effects. Then, our proposed STR, $\widehat{\delta}_{\drpt}$, is defined by substituting $\widehat{\beta}_t$ for $\beta_t$ in \cref{eq:welfare-maximizing-ITR-stated-preference}. For simplicity, we set $(\widehat{\delta}_{\drpt}(1),\widehat{\delta}_{\drpt}(0)) = (0,0)$ when $\widehat{\beta}_0 = \widehat{\beta}_1 = 0$ or when $\widehat{\beta}_1 < 0 < \widehat{\beta}_0$ and $\widehat{\beta}_1 + \widehat{\beta}_0 = 0$.
\par
The next result gives an upper bound of the finite-sample maximum regret of $\widehat{\delta}_{\drpt}$.
\begin{proposition}\label{proposition:maximum-regret-DRPT}
    Suppose that \cref{assumption:preference-for-lotteries,assumption:DRPT} hold. Then, for $n \geq \kappa^{-2}$,
    \begin{equation*}
        \sup_{P \in \mathcal{P}_{\drpt}(M,\kappa)} \mathbb{E}_{P^n}[\max_{\delta} W_{\stated}(\delta) - W_{\stated}(\widehat{\delta}_{\drpt})] \leq \frac{2e^{-1/2}M}{\kappa\sqrt{n}}.
    \end{equation*}
\end{proposition}
\cref{proposition:maximum-regret-DRPT} ensures that the maximum regret of $\widehat{\delta}_{\drpt}$ converges to zero at rate $n^{-1/2}$ as long as the data comes from the data generating process with \cref{assumption:DRPT}. This convergence rate is the same as that of $\widehat{\delta}_{\sprct}$ in \cref{proposition:maximum-regret-SP-RCT}.

\section{Empirical Application}
\label{sec:empirical-application}

We demonstrate our proposed STR using the results reported in \textcite{Wing2017}. They analyzed data from a DRPT conducted with students in an introductory psychology class \parencite{Clark2000}. This DRPT examined the effect of vocabulary and mathematics training on test scores. The total number of participants in this DRPT was 450, and they were randomly assigned to one of the three groups: the vocabulary training group, mathematics training group, and free-choice group with probability $1/4$, $1/4$, and $1/2$, respectively. As a result, three experimental groups, the vocabulary training group, mathematics training group, and free-choice group, contained 116, 119, and 210 students, respectively.  Fifty advanced vocabulary terms were taught in the vocabulary training group, while 5 algebraic concepts were taught in the mathematics training. In the following analysis, we regard vocabulary training as treatment 1 and mathematics training as treatment 0. Both treatments lasted about 15 minutes. After the training session, the participants took a post-test consisting of 30 vocabulary questions and 20 mathematics questions, regardless of which training was received. Of the 450 participants, 445 completed this experimental procedure. For a more detailed description of this experiment, see \textcite{Shadish2008}. 
\par
\cref{tab:emp-app}, adapted from \textcite{Wing2017}, shows the estimates of the share of the preferred treatment and the estimates of the CATEs on vocabulary and mathematics test scores given the preferred treatment. The estimates imply that $62\%$ of students prefer vocabulary training while $32\%$ prefer mathematics training. For students who preferred vocabulary learning, vocabulary learning improved vocabulary test scores by 8.5 points and reduced math test scores by 3.4 points compared to math learning. For students who preferred learning mathematics, vocabulary learning improved vocabulary test scores by 7.4 points and reduced mathematics test scores by 5.5 points compared to mathematics learning. All of the CATEs were significantly different from zero.
\begin{table}[t]
    \centering
    \begin{threeparttable}
        \caption{Share of Preferred Training and Conditional Average Treatment Effect}
        \label{tab:emp-app}
        \begin{tabular*}{\textwidth}{@{\extracolsep{\fill}}lccc}
            \toprule
            && \multicolumn{2}{c}{Conditional Average Treatment Effect}\\
            \cmidrule{3-4}
            Preferred Treatment & Share & Vocabulary Score & Mathematics Score\\
            \midrule
            Vocabulary Training &  $0.62$ & $8.5$ & $-3.4$ \\
            & & $(0.6)$ & $(0.6)$\\
            Mathematics Training & $0.38$ & $7.4$ & $-5.5$ \\
            & & $(1.1)$ & $(1.2)$\\
            \bottomrule
        \end{tabular*}
        {\footnotesize
            \begin{tablenotes}
                \item\textit{Notes}: This table is adapted from Tables~1 and 2 in \textcite[pp. 430 and 431]{Wing2017}. The table shows the estimates of the share of the preferred treatment and the conditional average treatment effect of vocabulary training relative to mathematics training on the test scores. The values in parentheses are the standard errors of the corresponding estimates.
            \end{tablenotes}
        }
    \end{threeparttable}
\end{table}
\par
For illustrational purposes, we define the outcome of interest as the weighted sum of vocabulary and mathematics test scores. Formally, let $V_i(d)$ and $M_i(d)$ be the potential vocabulary and mathematics test scores under treatment $d$. Given a weight $w \in [0,1]$, the potential outcome of interest under treatment $d$ is defined by
\begin{equation*}
    Y_i(d) = (1-w)V_i(d) + wM_i(d).
\end{equation*}
The weight being equal to zero means we only care about the vocabulary test scores. As $w$ gets large, more emphasis is put on the mathematics test scores, and $w=1$ means that we focus only on the mathematics test scores. With this definition of the targeted outcome, we operate the STR proposed in \cref{sec:doubly-randomized-preference-trials}.
\par
\begin{figure}[t]
    \centering
    \caption{Determinants of ITR}
    \includegraphics{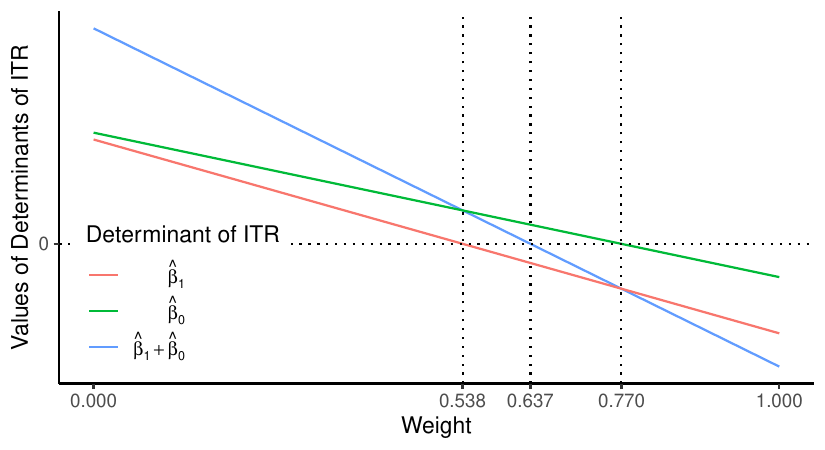}
    \label{fig:determinants-of-ITR}
\end{figure}

\cref{fig:determinants-of-ITR} draws determinants of ITR---$\widehat{\beta}_1$, $\widehat{\beta}_0$, and $\widehat{\beta}_1 + \widehat{\beta}_0$---by each weight of the targeted outcome. When the weight is less than $0.538$, both $\widehat{\beta}_1$ and $\widehat{\beta}_0$ are positive. Hence, our proposed STR indicates all students take the vocabulary training. When the weight is larger than $0.770$, both $\widehat{\beta}_1$ and $\widehat{\beta}_0$ is negative. In this case, the STR indicates all students take the mathematics training. When the weight is in $(0.538,0.770)$, $\widehat{\beta}_1$ is negative and $\widehat{\beta}_0$ is positive. At first glance, it would seem optimal to instruct those who prefer vocabulary training to learn math and those who prefer math training to learn vocabulary. However, upon learning of this ITR, students lie in their stated preferences. The resulting allocation achieved is not optimal. Instead, our STR does not personalize the assignment based on stated preferences but rather assigns the same training to all students. Specifically, when the weight is less than or equal to $0.637$, we assign vocabulary learning to all students; otherwise, we assign math learning to all students. 

\section{Conclusion}
\label{sec:conclusion}

This study investigated the individualized treatment assignment problem based on stated preferences for treatments. When individuals know the deployed ITR before the preference statement, they strategically state their preferences. Under the assumption that individuals maximize their expected utility, we derived an optimal ITR that maximizes welfare. The optimal ITR is strategy-proof, that is, individuals have no strong incentive to make a false preference statement. The optimal ITR requires information about the distribution of the true treatment preference and the conditional average treatment effect given the true preference. We proposed two experimental designs---strictly strategy-proof RCTs (SSP-RCTs) and doubly randomized preference trials (DRPTs)---that allow us to identify the information. We developed statistical treatment rules, assuming that the data comes from either SSP-RCTs or DRPTs. The maximum regret of the proposed STRs converges to zero at a rate of the square root of the sample size.
\par
We focused on binary treatment assignment problems and have not mentioned the case of more than three treatments. It is easy to adapt the model developed in \cref{sec:model} to accommodate more than three treatments. Here, we briefly demonstrate this for the case of three treatments. Each individual has a strict preference for the three treatments. Then, we can divide individuals into six types of preferences. Accordingly, the CATEs are defined for each preference type, and an ITR specifies a lottery over the treatments for each preference type. To extend the preference for treatments to the preference over lotteries, we can utilize the concept of first-order stochastic dominance \parencite[see, e.g.,][]{Erdil2014,Erdil2008,Kojima2010}. However, the form of the optimal ITR is unclear when three treatments exist. This is left for future research. 
\par
Another issue not addressed in this study is an ethical one. When people have a preference for a treatment, is it ethical to give them a treatment that differs from their preferred treatment? In the medical context, this may be permissible as the physician often has more knowledge about the treatment than the patient and may be able to persuade the patient to accept the recommended treatment. However, this is not always permissible in public policy, and the pros and cons may vary depending on the context.

\appendix
\section{Proofs}
\label{sec:proofs}
\begin{proof}[Proof of \cref{lemma:necessary-strategy-proofness}]
    Consider any individual who prefers treatment 1. Under \cref{assumption:preference-for-lotteries}, $\delta$ is [strictly] strategy-proof for this individual if and only if 
    \begin{equation*}
        \delta(1) \geq [>] ~ \delta(0).
    \end{equation*}
    Conversely, consider an individual whose true preference type is 0. Again, $\delta$ is [strictly] strategy-proof for this individual if and only if
    \begin{equation*}
        1 - \delta(0) \geq [>] ~ 1 - \delta(1).
    \end{equation*}
    Combining the observations made concludes the proof of the first statement. The second statement can be shown in a similar way.
\end{proof}
\begin{proof}[Proof of \cref{proposition:optimal-assignment-under-stated-preference}]
    For ease of exposition, we number the cases in the right-hand side of \cref{eq:welfare-maximizing-ITR-stated-preference} from top to bottom, which yields cases $1,\cdots,11$. \cref{lemma:necessary-strategy-proofness} readily implies that $\delta^*$ is strategy-proof in any case. The remaining task is to show that $\delta^*$ maximizes the welfare $W_{\stated}(\cdot)$ given in \cref{eq:def-welfare-stated-preference}. Here, we prove the welfare-maximizing property in cases 1 and 7. For the other cases, the property can be shown in a similar way as that of case 1 or 7.\par
    Case 1: $\mathbb{P}(T_i = 1) \tau(1) > 0$ and $\mathbb{P}(T_i = 0) \tau(0) > 0$. From \cref{eq:welfare-stated-preference-expansion}, observe that $W_{\stated}(\delta^*) = \mathbb{P}(T_i = 1) \tau(1) + \mathbb{P}(T_i = 0) \tau(0) + \mathbb{E}[Y_i(0)]$. Consider ITRs $\delta$ with $\delta(1) > \delta(0)$. Under \cref{assumption:preference-for-lotteries}, \cref{eq:welfare-stated-preference-expansion} implies that
    \begin{equation}
        W_{\stated}(\delta) = \mathbb{P}(T_i = 1) \tau(1) \delta(1) + \mathbb{P}(T_i = 0) \tau(0) \delta(0) + \mathbb{E}[Y_i(0)],
        \label{eq:welfare-p1>p0}
    \end{equation}
    which is less than $W_{\stated}(\delta^*)$ as long as $\delta(1) > \delta(0)$. Next, consider ITRs $\delta$ such that $\delta(1) < \delta(0)$. Given these ITRs, \cref{eq:welfare-stated-preference-expansion} gives
    \begin{equation}
        W_{\stated}(\delta) = \mathbb{P}(T_i = 1)  \tau(1)  \delta(0) + \mathbb{P}(T_i = 0)  \tau(0)  \delta(1) + \mathbb{E}[Y_i(0)].
        \label{eq:welfare-p1<p0}
    \end{equation}
    Again, the welfare is less than $W_{\stated}(\delta^*)$ as long as $\delta(1) < \delta(0)$. Finally, consider ITRs $\delta$ such that $\delta(1) = \delta(0) = \Bar{\delta}$. The welfare becomes 
    \begin{equation}
        W_{\stated}(\delta) = \mathbb{E}[Y_i(1) - Y_i(0)] \Bar{\delta} + \mathbb{E}[Y_i(0)],
        \label{eq:welfare-p1=p0}
    \end{equation}
    which is not greater than $W_{\stated}(\delta^*)$. Combining the observations made, we conclude that $\delta^*$ maximizes welfare in case 1.
    \par
    Case 7: $\mathbb{P}(T_i = 1) \tau(1) < 0$, $\mathbb{P}(T_i = 0) \tau(0) > 0$, and $\mathbb{E}[Y_i(1) - Y_i(0)] \leq 0$. Note that $W_{\stated}(\delta^*) = \mathbb{E}[Y_i(0)]$. Consider ITRs $\delta$ with $\delta(1) > \delta(0)$, whence \cref{eq:welfare-p1>p0} holds. For a fixed welfare level $\bar{W}$, consider the iso-welfare line of level $\bar{W}$ to be the collection of ITRs satisfying
    \begin{align*}
        &\bar{W} = \mathbb{P}(T_i = 1)  \tau(1)  \delta(1) + \mathbb{P}(T_i = 0)  \tau(0)  \delta(0) + \mathbb{E}[Y_i(0)].
        \label{eq:iso-welfare-p1>p0}\\
        \iff & \delta(0) = -\frac{\mathbb{P}(T_i=1) \tau(1)}{\mathbb{P}(T_i = 0)  \tau(0)} \cdot \delta(1) + \frac{\bar{W} - \mathbb{E}[Y_i(0)]}{\mathbb{P}(T_i=0)  \tau(0)}.\nonumber
    \end{align*}
    As $\mathbb{E}[Y_i(1) - Y_i(0)] > 0$, the iso-welfare line is steeper than the 45-degree line in the $x$-$y$ plane where $x$ and $y$ correspond to $\delta(1)$ and $\delta(0)$, respectively. Then, it can be easily confirmed that for any ITR satisfying $\delta(1) > \delta(0)$, its welfare is less than $\mathbb{E}[Y_i(0)]$. Next, consider ITRs $\delta$ such that $\delta(1) < \delta(0)$, under which \cref{eq:welfare-p1<p0} holds. In this case, the iso-welfare line of level $\bar{W}$ becomes
    \begin{equation*}
        \delta(0) = -\frac{\mathbb{P}(T_i=0) \tau(0)}{\mathbb{P}(T_i = 1) \tau(1)} \cdot \delta(1) + \frac{\bar{W} - \mathbb{E}[Y_i(0)]}{\mathbb{P}(T_i=1) \tau(1)}
    \end{equation*}
    As $\mathbb{E}[Y_i(1) - Y_i(0)] \leq 0$, the iso-welfare line is not steeper than the 45-degree line in the same $x$-$y$ plane as above. Then, the welfare of ITRs with $\delta(1) < \delta(0)$ is less than $W_{\stated}(\delta^*)$. Finally, consider ITRs such that $\delta(1) = \delta(0) = \Bar{\delta}$, whose welfare can be expressed by \cref{eq:welfare-p1=p0}. It is obvious that the welfare is not greater than $W_{\stated}(\delta^*)$ because $\mathbb{E}[Y_i(1) - Y_i(0)] \leq 0$. Combining the arguments made shows that $\delta^*$ is welfare-maximizing in case 7. 
\end{proof}
\begin{proof}[Proof of \cref{proposition:maximum-regret-SP-RCT}]
For notational simplicity, let $\beta_t = \mathbb{P}(T_i = t) \tau(t)$ for each $t \in \mathcal{T}$. Referring to \cref{tab:comparison-assignments}, we observe that $\delta^*$ takes a different form in the following three cases:
\begin{equation*}
    \delta^* = 
    \begin{cases}
        (1,0) & \text{if} \quad \beta_1 > 0 \land \beta_0 < 0,\\
        (1,1) & \text{if} \quad (\beta_1 \leq 0 \lor \beta_0 \geq 0) \land \beta_1 + \beta_0 > 0,\\
        (0,0) & \text{if} \quad (\beta_1 \leq 0 \lor \beta_0 \geq 0) \land \beta_1 + \beta_0 \leq 0.\\
    \end{cases}
\end{equation*}
Accordingly, the maximized welfare $W_{\stated}(\delta^*)$ can be written as
\begin{equation*}
    W_{\stated}(\delta^*) = 
    \begin{cases}
        \beta_1 + \mathbb{E}[Y_i(0)] & \text{if} \quad \beta_1 > 0 \land \beta_0 < 0,\\
        \beta_1 + \beta_0 + \mathbb{E}[Y_i(0)] & \text{if} \quad (\beta_1 \leq 0 \lor \beta_0 \geq 0) \land \beta_1 + \beta_0 > 0,\\
        \mathbb{E}[Y_i(0)] & \text{if} \quad (\beta_1 \leq 0 \lor \beta_0 \geq 0) \land \beta_1 + \beta_0 \leq 0.\\
    \end{cases}
\end{equation*}
Similarly, the STR $\widehat{\delta}_{\sprct}$ and its welfare $W_{\stated}(\widehat{\delta}_{\sprct})$ take a different form in the same cases except that $\beta_t$ is replaced with $\widehat{\beta}_t$.
\par
Consider joint distributions in $\mathcal{P}_{\sprct}(M,\kappa)$ such that $\beta_1 > 0$ and $\beta_0 < 0$. In this case, the regret of $\widehat{\delta}_{\sprct}$ can be bounded from above as follows:
\begin{align*}
    &\mathbb{E}_{P^n}[W_{\stated}(\delta^*) - W_{\stated}(\widehat{\delta}_{\sprct})]\\
    = {} & \{\beta_1 + \mathbb{E}[Y_i(0)] - (\beta_1 + \beta_0 + \mathbb{E}[Y_i(0)])\} \mathbb{P}\left((\widehat{\beta}_1 \leq 0 \lor \widehat{\beta}_0 \geq 0) \land \widehat{\beta}_1 + \widehat{\beta}_0 > 0\right)\\
    & + (\beta_1 + \mathbb{E}[Y_i(0)] - \mathbb{E}[Y_i(0)]) \mathbb{P}\left((\widehat{\beta}_1 \leq 0 \lor \widehat{\beta}_0 \geq 0) \land \widehat{\beta}_1 + \widehat{\beta}_0 \leq 0\right)\\
    \leq {} & \left(|\beta_1| + |\beta_0|\right) \left\{\mathbb{P}\left(\widehat{\beta}_1 \leq 0\right) + \mathbb{P}\left(\widehat{\beta}_0 \geq 0\right)\right\}.
\end{align*}
As $\beta_1 > 0$ by assumption, we have
\begin{align*}
    \mathbb{P}\left(\widehat{\beta}_1 \leq 0\right) = \mathbb{P}\left(\widehat{\beta}_1 - \beta_1 \leq - \beta_1\right).
\end{align*}
Note that $\widehat{\beta}_t$ is a sum of independent random variables, each of which is in $[-M/(n\kappa),M/(n\kappa)]$ by \cref{assumptionenum:bounded-outcome-SPRCT,assumptionenum:strict-overlap-SPRCT}. In addition, \cref{assumptionenum:unconfoundedness-SPRCT,assumptionenum:agreement-SPRCT} ensure that $\widehat{\beta}_t$ is an unbiased estimator for $\beta_t$. Hence, Hoeffding's inequality \parencite{Hoeffding1963} yields
\begin{align*}
    \mathbb{P}\left(\widehat{\beta}_1 \leq 0\right) &= \mathbb{P}\left(\widehat{\beta}_1 - \beta_1 \leq -\beta_1\right) \leq \exp\left(-\frac{\beta_1^2\kappa^2n}{2M^2}\right), \text{ and}\\
    \mathbb{P}\left(\widehat{\beta}_0 \geq 0\right) &= \mathbb{P}\left(\widehat{\beta}_0 - \beta_0 \geq -\beta_0\right) \leq \exp\left(-\frac{\beta_0^2\kappa^2n}{2M^2}\right).
\end{align*}
Then, the worst-case regret of $\widehat{\delta}_{\sprct}$ over the subclass of $\mathcal{P}_{\sprct}(M,\kappa)$ satisfying $\beta_1 > 0$ and $\beta_0 < 0$ can be bounded by
\begin{align*}
    \MoveEqLeft
    \max_{(\beta_1,\beta_0) \in [0,M]^2}\left[(\beta_1 + \beta_0)\left\{\exp\left(-\frac{\beta_1^2\kappa^2n}{2M^2}\right) + \exp\left(-\frac{\beta_0^2\kappa^2n}{2M^2}\right)\right\}\right]\\
    \leq {} & \max_{\beta_1 \in [0,M]}\left\{\beta_1\exp\left(-\frac{\beta_1^2\kappa^2n}{2M^2}\right)\right\} + \max_{\beta_0\in[0,M]}\left\{\beta_0\exp\left(-\frac{\beta_0^2\kappa^2n}{2M^2}\right)\right\}\\
    = {} & \frac{2e^{-1/2}M}{\kappa\sqrt{n}},
\end{align*}
where the last equality is obtained by solving the two maximization problems, given $n \geq \kappa^{-2}$.
\par
Next, consider joint distributions in $\mathcal{P}_{\sprct}(M,\kappa)$ such that $\beta_1 \leq 0$ or $\beta_0 \geq 0$ and that $\beta_1 + \beta_0 > 0 $. It follows that
\begin{align*}
    &\mathbb{E}_{P^n}[W_{\stated}(\delta^*) - W_{\stated}(\widehat{\delta}_{\sprct})]\\
    = {} &\{\beta_1 + \beta_0 + \mathbb{E}[Y_i(0)] - (\beta_1 + \mathbb{E}[Y_i(0)])\}\mathbb{P}\left(\widehat{\beta}_1 > 0 \land \widehat{\beta}_0 < 0\right)\\
    & + (\beta_1 + \beta_0 + \mathbb{E}[Y_i(0)] - \mathbb{E}[Y_i(0)])\mathbb{P}\left((\widehat{\beta}_1 \leq 0 \lor \widehat{\beta}_0 \geq 0) \land \widehat{\beta}_1 + \widehat{\beta}_0 \leq 0\right)\\
    \leq {} & \beta_0\mathbb{P}\left(\widehat{\beta}_0 \leq 0\right) + (\beta_1 + \beta_0)\mathbb{P}\left(\widehat{\beta}_1 + \widehat{\beta}_0 \leq 0\right).
\end{align*}
Again, Hoeffding's inequality guarantees that 
\begin{align*}
    \mathbb{P}\left(\widehat{\beta}_0 \leq 0\right) \leq \exp\left(-\frac{\beta_0^2\kappa^2n}{2M^2}\right) \text{ and }
    \mathbb{P}\left(\widehat{\beta}_1 + \widehat{\beta}_0 \leq 0\right) \leq \exp\left(-\frac{(\beta_1 + \beta_0)^2\kappa^2n}{2M^2}\right).
\end{align*}
Given these inequalities, the upper bound of the maximum regret over distributions in $\mathcal{P}_{\sprct}(M,\kappa)$ satisfying $(\beta_1 \leq 0 \lor \beta_0 \geq 0) \land \beta_1 + \beta_0 > 0$ is bounded by
\begin{align*}
    \MoveEqLeft
    \max_{(\beta_1,\beta_0) \in [0,M]^2}\left\{\beta_0\exp\left(-\frac{\beta_0^2\kappa^2n}{2M^2}\right) + (\beta_1 + \beta_0)\exp\left(-\frac{(\beta_1+\beta_0)^2\kappa^2n}{2M^2}\right)\right\}\\
    \leq {} & \max_{\beta_0\in[0,M]}\left\{\beta_0\exp\left(-\frac{\beta_0^2\kappa^2n}{2M^2}\right)\right\} + \max_{\beta \in [0,2M]}\left\{\beta\exp\left(-\frac{\beta^2\kappa^2n}{2M^2}\right)\right\}\\
    = {} & \frac{2e^{-1/2}M}{\kappa\sqrt{n}}.
\end{align*}
\par
The upper bound of the worst-case regret over the remaining class of joint distributions can be obtained almost in the same way as the preceding paragraph. Combining the results presented concludes the proof.
\end{proof}
\begin{proof}[Proof of \cref{proposition:maximum-regret-DRPT}]
    Observe that $\widehat{\beta}_t$ is a sum of iid random variables, whose range is in $[-M/(n\kappa),M/(n\kappa)]$ by \cref{assumptionenum:bounded-outcome-DRPT,assumptionenum:strict-overlap-DRPT}. In addition, $\widehat{\beta}_t$ is an unbiased estimator for $\beta_t$ given \cref{assumptionenum:random-assignment-DRPT,assumptionenum:existence-exogeneous-free-choice,assumptionenum:exclusion-restriction-DRPT}. Then, the desired result can be shown in an identical way as that of \cref{proposition:maximum-regret-SP-RCT}. 
\end{proof}

\printbibliography

\end{document}